\newif\ifanonymoussubmission
\anonymoussubmissionfalse

\documentclass[letterpaper]{article} % DO NOT CHANGE THIS
\ifanonymoussubmission
  \usepackage[submission]{aaai2027} % DO NOT CHANGE THIS
\else
  \usepackage[preprint]{aaai2027} % DO NOT CHANGE THIS
\fi
\usepackage[hyphens]{url} % DO NOT CHANGE THIS
\usepackage{graphicx} % DO NOT CHANGE THIS
\usepackage{natbib} % DO NOT CHANGE THIS
\usepackage{caption} % DO NOT CHANGE THIS
\usepackage{amsmath,amssymb,amsthm,bm}
\usepackage{booktabs}
\usepackage{algorithm}
\usepackage{algpseudocode}
\usepackage{tikz}
\usepackage{multirow,color}
\usetikzlibrary{arrows.meta,positioning}

\pdftrailerid{}
\newtheorem{theorem}{Theorem}

\theoremstyle{definition}
\newtheorem{definition}{Definition}
\newtheorem{assumption}{Assumption}
\theoremstyle{remark}

\newcommand{\V}{\mathcal V}
\newcommand{\Key}{\mathsf{Key}}
\newcommand{\Unif}{\mathsf{Unif}}
\newcommand{\indep}{\perp\!\!\!\perp}

\newcommand{\cP}{\mathcal P}

\title{%PIVOT: Model-Free Multi-Bit LLM Watermarking with Mirrored Rank-Interval Sampling
PIVOT: Distribution-Preserving Multi-Bit Watermarking with Minimax-Optimal Scoring
}
\ifanonymoussubmission
\author{Anonymous Submission}
\affiliations{}
\else
\author{
Wen Zhang\textsuperscript{\rm 1},
Kaiyi Pang\textsuperscript{\rm 2},
Minhao Bai\textsuperscript{\rm 2},
Shuyu Gan\textsuperscript{\rm 3},
Liping Zhu\textsuperscript{\rm 4}\corresponding
}
\affiliations{
\textsuperscript{\rm 1}Institute of Statistics and Big Data,
Renmin University of China\\
\textsuperscript{\rm 2}Department of Electronic Engineering,
Tsinghua University\\
\textsuperscript{\rm 3}Department of Mathematical Sciences,
Tsinghua University\\
\textsuperscript{\rm 4}Centre for Applied Statistics and the
Institute of Statistics and Big Data,
Renmin University of China\\
zhangwen.stat@gmail.com,
zhu.liping@ruc.edu.cn\\}
\fi

\begin{document}
\maketitle

\begin{abstract}
As LLM-generated content becomes increasingly widespread, reliable attribution calls for multi-bit watermarks that embed recoverable messages without distorting the output distribution.
We introduce
PIVOT, a multi-bit watermark with model-free verification based on mirrored rank-interval sampling. At each generation
step, PIVOT uses a keyed pseudorandom scalar and a secret permutation of the vocabulary. The message
bit determines whether next-token probability intervals are arranged in forward or reverse secret-rank
order. Sampling from these mirrored layouts preserves every token probability exactly while inducing
an opposite-signed dependence between the scalar and the sampled token's secret rank. This dependence
yields a pivotal rank statistic whose null distribution is invariant to the unknown next-token law,
enabling model-free calibration and decoding. We formulate joint watermark detection and message
recovery as a multiple-hypothesis testing problem, derive a moment-robust minimax-optimal detection
score over a relaxed moment class, and establish finite-sample recovery guarantees.
Extensive experiments on two representative large language models validate the effectiveness of PIVOT, demonstrating reliable watermark detection and message recovery
while preserving the generator's output distribution.
\end{abstract}

\section{Introduction}

As LLMs increasingly generate high-impact content, reliable provenance is essential for auditing misuse and attributing model outputs \cite{liu2023survey}. Passive AI-text detectors can be brittle under distribution shift and paraphrasing \cite{sadasivan2023reliably}; generative watermarking instead embeds a keyed statistical signal during sampling. Early LLM watermarks were primarily zero-bit and supported only binary detection. Although subsequent work improved calibration, scalability, security, and distribution preservation \cite{kirchenbauer2023watermark,hu2023unbiased,dathathri2024synthid,christ2024undetectable,wu2024dipmark,zhao2025pf}, a zero-bit flag cannot identify the model, user, account, or generation event responsible for the text.

% muti-bit
%Multi-bit watermarking extends provenance from identification to attribution by embedding a recoverable message, such as a model identifier, user or tenant ID, timestamp, or database index. Early codable schemes generalized vocabulary partitioning to multiple symbols \cite{wang2023codable}. Modular constructions combined several watermarking primitives to support payload recovery \cite{fernandez2023threebricks}, while position-allocation methods divided a text among message blocks to support longer and more corruption-tolerant messages \cite{yoo2024multibit}. Provably robust multi-bit designs have also studied pseudo-random segment assignment and edit tolerance \cite{qu2024provably}. Beyond token-level generation, multi-bit and provenance-oriented watermarking has been explored for invariant linguistic features and scalable text attribution \cite{yoo2023invariant,lau2024waterfall}. Recent LLM-specific methods investigate multilayer unbiased reweighting \cite{feng2025bimark}, message-symmetric probability partitions \cite{zhu2026quantilemark}, short-text recovery \cite{xu2026xmark}, and stealthy embedding \cite{jiang2025stealthink}. Despite this progress, multi-bit watermarking creates a substantially harder design problem than zero-bit detection: a useful scheme must preserve generation quality, control false alarms, recover one message among many, remain computationally practical, and extract sufficient evidence from a limited number of tokens.

Multi-bit watermarking addresses this limitation by embedding a recoverable payload, such as a model identifier, user ID, or timestamp. Existing methods have increased capacity through codable vocabulary partitions and modular constructions \cite{wang2023codable,fernandez2023threebricks}, improved robustness through block or segment allocation \cite{yoo2024multibit,qu2024provably}, and explored invariant features, scalable attribution, unbiased reweighting, and short-text recovery \cite{yoo2023invariant,lau2024waterfall,feng2025bimark,zhu2026quantilemark,xu2026xmark,jiang2025stealthink}. Nevertheless, multi-bit watermarking must simultaneously preserve text quality, control false alarms, and reliably distinguish the true payload from many competitors.

Among these requirements, distribution preservation is particularly important. Although many of these methods achieve strong detection or recovery performance, they often obtain the watermark signal by perturbing the next-token law. Such perturbations can degrade generation quality and leave a payload-dependent distributional footprint \cite{kirchenbauer2023watermark,yoo2024multibit,hou2024semstamp}. Average or expectation-level preservation mitigates this issue but does not ensure that every fixed payload and history induce the original LLM distribution \cite{hu2023unbiased,feng2025bimark}. Distribution-preserving methods such as DiPmark and Gumbel-max sampling show that the signal can instead be placed in a keyed coupling \cite{wu2024dipmark,fernandez2023threebricks}. For multi-bit attribution, we seek the stronger message-conditional guarantee: every payload should induce exactly the same text distribution.

Preserving the generator distribution resolves the encoding-side challenge, but it does not determine how the detector should aggregate weak token-level evidence. \citet{listatistical} developed a principled statistical framework for zero-bit watermarks, combining pivotal statistics, distribution-free Type-I control, and minimax score design. Multi-bit verification is qualitatively different because an error can arise either from missing the watermark or from confusing the true payload with a competitor. Existing information-theoretic limits characterize general trade-offs \cite{he2026tradeoffs}, but do not provide a model-free statistic or an implementable score for a concrete multi-bit channel.

%Motivated by these gaps, we introduce PIVOT, a distribution-preserving multi-bit watermark with moment-robust detection. First, mirrored rank-interval sampling preserves every next-token probability for each fixed payload and history, while inducing a keyed dependence that yields a model-free pivotal statistic. Second, we formulate joint detection and recovery as a multi-hypothesis testing problem and derive a minimax-optimal score within a relaxed moment class. Finally, we establish finite-sample detection and recovery guarantees and empirically evaluate the resulting method across models and payload lengths.

%{\blue Motivated by these gaps}, 
To address the coupled challenges of message-conditional distribution preservation
and model-free multi-bit detection,
we introduce PIVOT (Permutation-Interval Variable-Order Text Watermarking), a distribution-preserving
multi-bit watermark equipped with a moment-robust minimax-optimal detection score.
At each generation step, the secret key produces a scalar and a random
permutation of the vocabulary. The next-token probabilities are represented by intervals of exactly
the same lengths, while the embedded bit determines whether these intervals are arranged in forward
or reverse secret-rank order. Changing the order alters the keyed dependence between the scalar and
the sampled token's rank, but leaves every token probability unchanged. The verifier reconstructs
this dependence from the key, converts each token into model-free pivotal evidence, and aggregates
the evidence blockwise to detect the watermark and recover the payload. We further design the local
score through a robust optimization problem rather than selecting it heuristically.

Our contributions are threefold.
\begin{itemize}
\item[(i)] We establish exact message-conditional distribution preservation and construct a pivotal
statistic whose null law does not depend on the unknown language-model probabilities, enabling
calibration without model access.
\item[(ii)] We formulate joint detection and recovery as a multiple-hypothesis testing problem and
derive a moment-robust minimax-optimal score for the detection component.
\item[(iii)] We provide finite-sample guarantees for complete-message recovery and evaluate PIVOT across
language models, payload lengths, and decoding settings. %including comparisons of detection, recovery, computational cost, and generation quality.
\end{itemize}

\section{Methodology}
\subsection{Problem setup}
We first formalize the multi-bit watermarking problem.
Let $\V$ be a finite vocabulary of size $V$. Given a prompt and a token history
$H_t=(w_1,\ldots,w_{t-1})$, an autoregressive language model specifies a next-token distribution
\[
  P_t=(p_t(v))_{v\in\V},\qquad
  \sum_{v\in\V}p_t(v)=1.
\]
Let $W^T=(W_1,\ldots,W_T)$ be an unwatermarked length-$T$ sequence, with joint distribution $Q_{W^T}$.
No independence, stationarity, or Markov assumption is imposed on this law unless stated otherwise.

An $L$-bit payload is indexed by a fixed $r\in[m]$, where
$m=2^L$ and $[m]=\{1,\ldots,m\}$.  Fix a bijection
$c:[m]\to\{0,1\}^L$ and write $c(r)=(c_1(r),\ldots,c_L(r))$.
The $T$ token positions are partitioned into fixed, public blocks
$\mathcal I_1,\ldots,\mathcal I_L$, with block $\mathcal I_\ell$ carrying bit $c_\ell(r)$.
Given the generated text and the secret key, the verifier must decide whether a watermark is
present and, if so, recover $r$.  We develop a
rank-based scheme for this joint detection-and-recovery task.
%Fixed blocks are essential for the information-theoretic statements below; message-dependent stopping is discussed separately in Section~\ref{sec:discussion}.
%We designed a rank-based multi-bit watermark and selected an optimal score function to maximize its detection efficiency.
  \begin{figure}[ht]
    \centering
    \includegraphics[
      width=\columnwidth,
      keepaspectratio
    ]{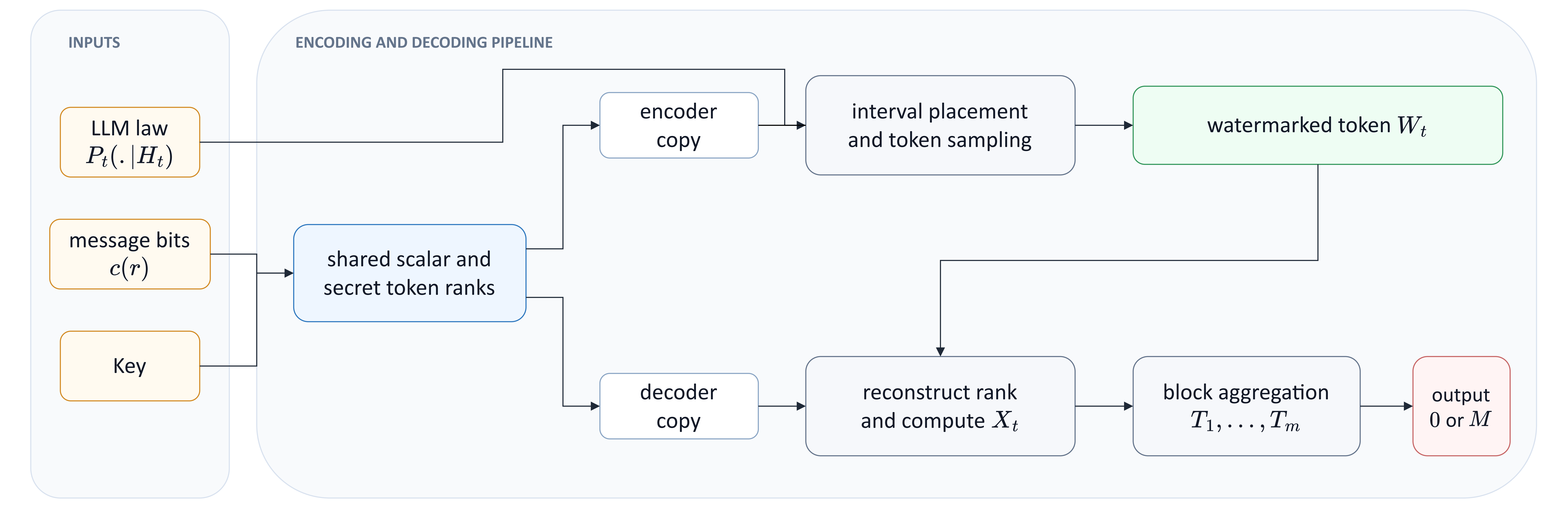}
    \caption{Encoding couples the message with model sampling through
  secret interval placement. Decoding reconstructs the same ranks, converts each token to pivotal
  evidence, and jointly tests all candidate messages.}
  \label{fig:pipeline}
  \end{figure}
\subsection{Encoding: coupling a message with model sampling}
A watermark typically consists of two parts: pseudorandom variables derived from a key and a
deterministic sampling function. We describe our construction through these two components.
\paragraph{Pseudorandom variables.}
The encoder and the authorized verifier share a secret key $\Key$. The key determines a
reproducible sequence
\begin{align*}
   \zeta_t=(U_t,\Pi_t)=G_{\Key}(\ell,i),
  \qquad t\in\mathcal I_\ell,
\end{align*}
where $i$ is the position within block $\mathcal I_\ell$,
$U_t\in[0,1)$ is a scalar, and
$\Pi_t:\V\to\{0,\ldots,V-1\}$ is a one-to-one assignment of secret ranks to tokens.
The detector can reproduce $(U_t,\Pi_t)$ from the key and public position, whereas an observer
without the key cannot determine either quantity.
%The statistical analysis does not depend on a particular cryptographic primitive. It starts from the ideal law
Under the idealized random-key model,
\begin{align}
  U_t\stackrel{\mathrm{iid}}{\sim}\Unif[0,1),\qquad
  \Pi_t\stackrel{\mathrm{iid}}{\sim}\Unif(\mathfrak S_V),\qquad
  U_t\indep\Pi_t.
  \label{eq:ideal-randomness}
\end{align}
where $\mathfrak S_V$ is the set of all permutations of the vocabulary $\V$ and the probability
is over the randomness of the key. A practical keyed generator is required only to make its output
indistinguishable from this ideal law to parties without $\Key$.

\begin{figure}[htbp]
  \centering
  \includegraphics[
    width=\columnwidth,
    keepaspectratio
  ]{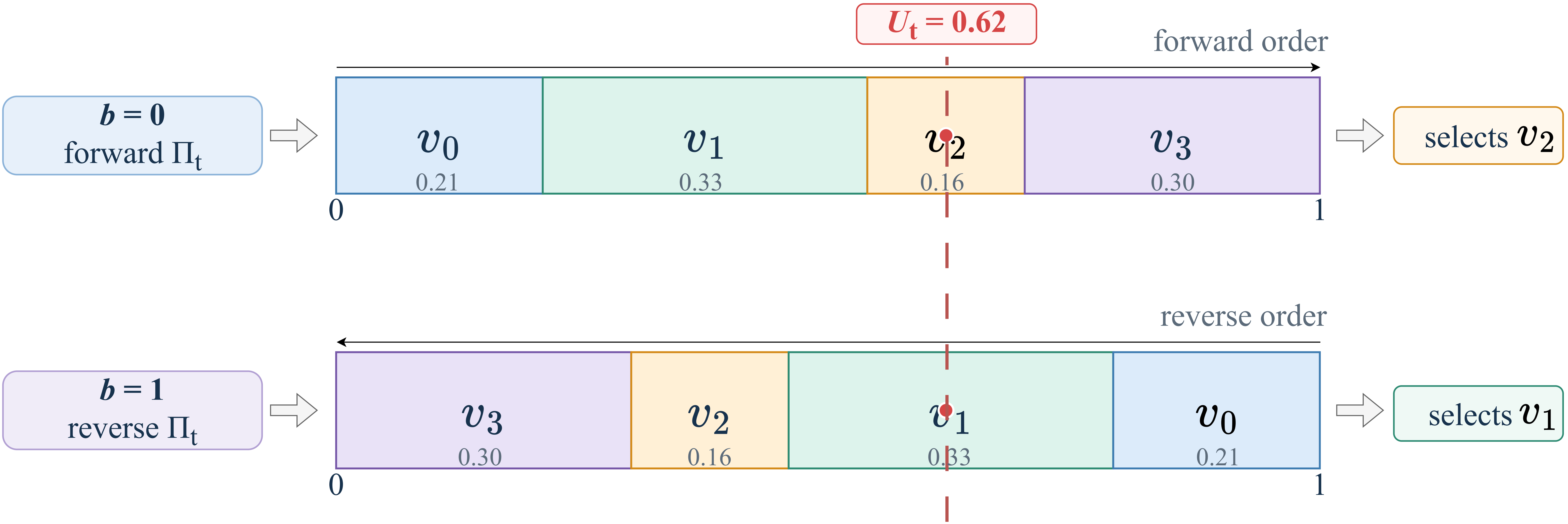}
  \caption{The same probability masses arranged in forward and reverse secret-rank order. The
  dashed scalar selects different ranks under the two embedded bits.}
  \label{fig:interval-reversal}
\end{figure}

\paragraph{Sampling method.}
%\paragraph{Probability intervals in secret-rank order.}
Figure~\ref{fig:interval-reversal} illustrates our sampling method. Each rectangle represents one
token, and its width equals that token's next-token probability. The permutation $\Pi_t$ assigns a
secret rank to every token. For
$b=0$, the encoder places these probability intervals from left to right in increasing secret-rank
order; for $b=1$, it places the same intervals in decreasing order. The shared scalar $U_t$ is
located on $[0,1)$, and the interval containing it determines the selected rank.  Reversing the order can change the
selected token for the same $U_t$, but it does not change any interval length and therefore
preserves every token's marginal probability.

We express this process mathematically.
At position $t$, place the model probabilities in increasing secret-rank order:
\begin{align}
  q_{t,r}
  &=p_t\!\left(\Pi_t^{-1}(r)\right),\notag\\
  F_{t,r}
  &=\sum_{a=0}^{r-1}q_{t,a},
  \qquad r=0,\ldots,V,
  \label{eq:rank-cdf}
\end{align}
with $F_{t,0}=0$ and $F_{t,V}=1$. Thus $q_{t,r}$ is the probability of the token having
secret rank $r$, and $[F_{t,r},F_{t,r+1})$ is an interval of exactly that length.
The bit $b\in\{0,1\}$ determines where this interval is placed:
\begin{align*}
    I_{t,r}^{(0)}
  =[F_{t,r},F_{t,r+1}),
  \qquad
  I_{t,r}^{(1)}
  =[1-F_{t,r+1},1-F_{t,r}).
\end{align*}
The second collection is the mirror image of the first around $1/2$. The encoder selects the unique
rank $R_t^{(b)}$ and emits the next token $W_t$ as
\begin{align}
  R_t^{(b)}=r
  &\quad\Longleftrightarrow\quad
  U_t\in I_{t,r}^{(b)},
  \quad r=0,\ldots,V-1,\notag\\
  W_t
  &=\Pi_t^{-1}\!\left(R_t^{(b)}\right).
  \label{eq:bit-sampler}
\end{align}
% 涓嶆槸闅忔満鍙橀噺搴旇鐢ㄥ皬鍐?w
The message changes only the locations of the intervals. Their lengths, and hence all token
probabilities, remain unchanged. The complete procedure is summarized in
Algorithm~\ref{alg:encoder}.
\begin{algorithm}[t]
  \caption{Fixed-block rank-based encoder}
  \label{alg:encoder}
  \begin{algorithmic}[1]
  \Require message $r$, blocks $(\mathcal I_\ell)_{\ell=1}^L$, language model, secret key $\Key$
  \State compute the bit vector $c(r)=(c_1(r),\ldots,c_L(r))$
  \For{$\ell=1,\ldots,L$}
    \For{$t\in\mathcal I_\ell$}
      \State obtain $P_t(\cdot)=P(\cdot\mid H_t)$ from the language model
      \State reproduce $(U_t,\Pi_t)=G_{\Key}(\ell,i)$
      \State compute $(q_{t,r})_{r=0}^{V-1}$ and $(F_{t,r})_{r=0}^{V}$
      from~\eqref{eq:rank-cdf}
      \State find the unique $R_t$ such that
      $U_t\in I_{t,R_t}^{(c_\ell(r))}$
      \State emit $W_t=\Pi_t^{-1}(R_t)$ and append it to $H_t$
    \EndFor
  \EndFor
  \State \Return $W^T$
  \end{algorithmic}
  \end{algorithm}

To maintain output quality and prevent payload leakage through the text marginal, a watermarked LLM
should preserve the original output distribution, a property often called watermark unbiasedness.
The following theorem establishes this property for PIVOT.
\begin{theorem}[Distribution preservation]
\label{thm:unbiased}
For every distribution $P$ on $\V$, every $v\in\V$, and each $b\in\{0,1\}$,
\[
  \Pr\{W=v\mid b\}=p(v)
\]
under the encoding rule~\eqref{eq:bit-sampler}. Consequently, for every fixed message $r\in[m]$,
the induced distribution of $W^T$ equals $Q_{W^T}$.
\end{theorem}

The theorem is stronger than equality after averaging over messages: every message induces the
same observable text law as ordinary generation. The watermark is therefore carried entirely by
the coupling between $W^T$ and the secret auxiliary sequence
$\zeta^T=(\zeta_1,\ldots,\zeta_T)$.
Under fixed blocks, message-independent seed generation, and the ideal
experiment~\eqref{eq:ideal-randomness}, every payload induces the same marginal law for $W^T$
by Theorem~\ref{thm:unbiased}, while the law of $\zeta^T$ is message-independent by construction.
Here the probability is taken over fresh auxiliary randomness generated from an independently
sampled secret seed for each document.  The statement is therefore marginal over that seed, rather
than conditional on a realized key, for which $(U_t,\Pi_t)$ are deterministic.  %Reusing one long-term key across documents would require additional per-document randomization, such as a fresh nonce, and is not covered by the present theorem.

%Thus neither the text nor the secret sequence alone reveals the payload; the watermark resides only in their joint keyed dependence.

\subsection{Decoding: pivotal ranks and message scores}
We now turn to decoding: determining whether a candidate text carries the watermark and, if so,
recovering the embedded message. In realistic settings, the detector may lack access to the language
model, its parameters, or the original prompt and therefore cannot reconstruct the next-token
distribution $P_t$. This precludes the direct use of likelihood-ratio tests that depend on $P_t$.
We instead base detection and decoding on a pivotal statistic whose null distribution is invariant
to $P_t$, together with score functions constructed from that pivot.
\paragraph{A model-free pivotal statistic.}
Given a candidate text and the secret key, the verifier reconstructs $(U_t,\Pi_t)$ and computes
\[
  R_t=\Pi_t(W_t)\in\{0,\ldots,V-1\}.
\]
The verifier does not evaluate the language model and does not need $P_t$. Instead it forms
\begin{align}
  X_t
  =
  (2U_t-1)
  \left(\frac{2R_t}{V-1}-1\right)
  \in[-1,1].
  \label{eq:pivot}
\end{align}
Under an unwatermarked source, $W_t$ is unrelated to the current secret permutation and scalar.
Consequently, conditional on all previous observations, $R_t\sim\Unif\{0,\ldots,V-1\}$,
$R_t\indep U_t$, and
\begin{align}
  &\mathbb E_0[X_t\mid\mathcal F_{t-1}]=0,\quad \mathbb E_0[X_t^2\mid\mathcal F_{t-1}]
  =v_0:=\frac{V+1}{9(V-1)}.
  \label{eq:null-moments}
\end{align}
This null behavior is independent of the unknown language-model probabilities, which makes $X_t$
a pivot for false-alarm calibration.

Under watermarked sampling, the output rank is selected by an interval containing $U_t$.
Let $\sigma(0)=1$ and $\sigma(1)=-1$. If
$g(P)=1-\sum_{v\in\V}p(v)^2$ denotes the Gini impurity, then
\begin{align}
   \mathbb E\!\left[\sigma(b)X_t\mid P_t,b\right]
  =
  \kappa_V g(P_t),
  \qquad
  \kappa_V=\frac{V+1}{3(V-1)}.
  \label{eq:rank-drift}
\end{align}
Forward traversal creates positive drift and reverse traversal creates the reflected negative drift.
Nearly deterministic next-token distributions necessarily provide little signal.

%\paragraph{From local evidence to a complete-message decision.}
\paragraph{Score function.}
Let $h_0,h_1:[-1,1]\to\mathbb R$ be two local score functions. For a candidate message
$r\in[m]$, aggregate the score matching its bit in each block:
\begin{align}
    T_r(W^T,\zeta^T)
  =
  \sum_{\ell=1}^L
  \sum_{t\in\mathcal I_\ell}
  h_{c_\ell(r)}(X_t).
  \label{eq:generic-message-score}
\end{align}
Many score pairs are possible; the simplest is $h_0(x)=x$ and $h_1(x)=-x$.
Section~\ref{sec:moment-robust-score} derives a moment-robust minimax-optimal score.
Given a threshold $\tau_\alpha$ calibrated under unwatermarked generation, the detector outputs
\begin{align}
  \widehat r
  =
  \begin{cases}
    \displaystyle\arg\max_{r\in[m]}T_r,
      &\displaystyle\max_{r\in[m]}T_r>\tau_\alpha,\\[1mm]
    0,&\text{otherwise}.
  \end{cases}
  \label{eq:generic-decoder}
\end{align}
% 是否与逐步解码矛盾或统一
Thus detection and message recovery are performed jointly: the maximum score determines whether a watermark is present, while its maximizing index determines the decoded payload. 
In practice, the maximization need not enumerate all $m=2^L$ messages.  Define the two block scores
\begin{equation}
  S_{\ell,b}:=\sum_{t\in\mathcal I_\ell}h_b(X_t),
  \qquad
  \widehat b_\ell\in\arg\max_{b\in\{0,1\}}S_{\ell,b},
  \label{eq:bitwise-block-score}
\end{equation}
using a fixed rule to break ties.  Because $c$ is a bijection from $[m]$ onto $\{0,1\}^L$,
\begin{equation}
  \widetilde r
  :=c^{-1}(\widehat b_1,\ldots,\widehat b_L),
  \qquad
  T_{\widetilde r}
  =\sum_{\ell=1}^L S_{\ell,\widehat b_\ell}
  =\max_{r\in[m]}T_r.
  \label{eq:bitwise-joint-equivalence}
\end{equation}
Hence~\eqref{eq:generic-decoder} can be implemented by decoding each bit separately and then applying
the single global threshold $\tau_\alpha$ to the resulting maximum message score.  This equivalence
uses the full codebook $\{0,1\}^L$.

\begin{algorithm}[h]
\caption{Bitwise implementation of the fixed-block decoder}
\label{alg:decoder}
\begin{algorithmic}[1]
\Require token sequence $W^T$, blocks $(\mathcal I_\ell)$, bijection $c$, secret key $\Key$,
scores $(h_0,h_1)$, threshold $\tau_\alpha$
\For{$\ell=1,\ldots,L$}
  \For{the $i$-th position $t\in\mathcal I_\ell$}
    \State reproduce $(U_t,\Pi_t)=G_{\Key}(\ell,i)$
    \State compute $R_t=\Pi_t(W_t)$ and $X_t$ from~\eqref{eq:pivot}
  \EndFor
\EndFor
\For{$\ell=1,\ldots,L$}
  \State compute $S_{\ell,0}$ and $S_{\ell,1}$ from~\eqref{eq:bitwise-block-score}
  \State $\widehat b_\ell\gets\arg\max_{b\in\{0,1\}}S_{\ell,b}$
\EndFor
\State $\widetilde r\gets c^{-1}(\widehat b_1,\ldots,\widehat b_L)$
\State $T_{\max}\gets\sum_{\ell=1}^L S_{\ell,\widehat b_\ell}$
\If{$T_{\max}>\tau_\alpha$}
  \State \Return $\widetilde r$
\Else
  \State \Return $0$ \Comment{no watermark}
\EndIf
\end{algorithmic}
\end{algorithm}
\paragraph{Example.}
For the linear pair $h_0(x)=x$ and $h_1(x)=-x$, let
$Z_\ell=\sum_{t\in\mathcal I_\ell}X_t$.  Then
$S_{\ell,0}=Z_\ell$, $S_{\ell,1}=-Z_\ell$, and
\begin{equation*}
  \widehat b_\ell
  =
  \begin{cases}
    0,& Z_\ell\ge0,\\
    1,& Z_\ell<0,
  \end{cases}
\end{equation*}
where the first case fixes the tie-breaking rule.  Thus
$\widetilde r=c^{-1}(\widehat b_1,\ldots,\widehat b_L)$ and
$T_{\max}=\sum_{\ell=1}^L|Z_\ell|$.  The decoder returns $\widetilde r$ when
$T_{\max}>\tau_\alpha$ and returns $0$ otherwise.

\iffalse
For the complete-message statistic~\eqref{eq:generic-message-score}, define
\begin{equation*}
  \tau_\alpha
  :=
  \inf\left\{
    \tau:
    \sup_Q
    P_{Q\otimes P_{\zeta^T}}
    \left(
      \max_{r\in[m]}T_r>\tau
    \right)
    \leq\alpha
  \right\}.
\end{equation*}
After fixing a deterministic tie-breaking rule, the decoding regions are
\begin{equation*}
  \mathcal R_r
  =
  \left\{
    o:
    T_r(o)>\tau_\alpha,\quad
    r=\arg\max_{s\in[m]}T_s(o)
  \right\},
  \qquad r\in[m],
\end{equation*}
and the null acceptance region is
\[
  \mathcal A_0
  =
  \left\{
    o:\max_{r\in[m]}T_r(o)\leq\tau_\alpha
  \right\}.
\]
These regions partition the observation space and satisfy
$\sup_QP_0(\cup_{r=1}^m\mathcal R_r)\leq\alpha$ by construction. %Section~4 shows that choosing $T_r=\log L_r$ makes this size-constrained partition optimal for average message error.
\fi

\section{Statistical Framework}
This section develops a statistical framework for multi-bit watermarks based on
multiple-hypothesis testing, extending the conventional binary framework for zero-bit watermark
detection.
% 是否需要介绍notation
\subsection{Multiple-Hypothesis Testing}
Unlike zero-bit detection, which tests only whether a watermark is present,
multi-bit verification must also identify the embedded payload. We therefore
formulate joint detection and recovery as the following $(m+1)$-hypothesis test: %We therefore consider the following multiple-hypothesis test:
% Detection and decoding form the$(m+1)$-ary experiment\[\begin{array}{ll}H_0:&O\sim P_0:=Q_{W^T}\otimes P_{\zeta^T},\\[1mm]H_r,\ r\in[m]:&O\sim P_r:=P_{W^T,\zeta^T\mid M=r}.\end{array}\]
\[
\begin{array}{@{}l@{\;}l@{}}
H_0: & W^T\text{ is unwatermarked},\\[1mm]
H_r,\ r\in[m]: & W^T\text{ is watermarked with message }r,
\end{array}
\]
where $H_0$ denotes ordinary generation and $H_r$ denotes generation with message $r$.
%The null is a product law, whereas every alternative preserves both marginals and changes only their dependence.
The detector observes $O=(W^T,\zeta^T)$ and uses $\delta_r(o)$ to denote the probability of
outputting $r$ given $o$, where $\delta_r(o)\ge0$ and $\sum_{r=0}^m\delta_r(o)=1$. Define
\begin{align*}
  \beta_0(\delta)&=\mathbb E_0[1-\delta_0(O)],\\
  \beta_r(\delta)&=\mathbb E_r[1-\delta_r(O)],\qquad r\in[m].
\end{align*}
% $\Gamma=0$,$\Gamma\in[m]\setminus\{r\}$
For $r\in[m]$, $\beta_r$ counts both missed detections (output $0$) and incorrect message
recovery (output another message). This complete-message criterion differs from the bit error rate:
a single incorrect bit makes the entire message incorrect. In particular, when $m=1$, the
multiple-hypothesis test reduces to a standard binary test, and $\beta_0$ and $\beta_1$ are the
Type-I and Type-II errors, respectively. We define the worst-case modified Type-II error for
complete-message recovery as
%complete-message criteria
\begin{equation}
%\overline\beta(\delta)=\frac1m\sum_{r=1}^m\beta_r(\delta),\qquad
  \beta_{\max}(\delta)
  =\max_{r\in[m]}\beta_r(\delta).
  \label{eq:average-max}
\end{equation}
Our goal is to minimize $\beta_{\max}$ while controlling $\beta_0$.
We restrict attention to decoders induced by score functions $h$ and hence write
$\beta_0(h)$, $\beta_r(h)$, and $\beta_{\max}(h)$.
\begin{definition}[Size-constrained score function class]
\[
  \mathcal D_\alpha
  =
  \left\{
    h:
     \beta_0(h)\le\alpha
  \right\}.
\]
\end{definition}

%When robustness to an unknown null source is required, $\beta_0\le\alpha$ is replaced by $\blue \sup_Q\beta_0(h,Q\otimes P_{\zeta^T})\le\alpha$. 
When the null source is unknown, we require uniform size control: $\beta_0(h)$ must not exceed $\alpha$ for any admissible
null source.
The rank pivot permits such
distribution-free calibration.
%When the null source is unknown, we instead require $\beta_0(h;Q)\le\alpha$ for every admissible null law $Q$.
%\subsection{Null pivotality}
%Following the pivotal-statistic viewpoint of~\cite{listatistical}, detection is calibrated from a statistic $X_t$ whose null law is independent of the unknown language-model distribution. Under $H_0$, $W_t$ is independent of the fresh secret permutation and scalar. 

For any fixed measurable score pair $(h_0,h_1)$, the null distribution of the message statistics
in~\eqref{eq:generic-message-score} does not depend on the next-token distributions; hence, its
threshold can be calibrated without access to the language model. The resulting robust objective is
\begin{align*}
  %\inf_{\delta\in\mathcal D_\alpha^{\rm piv}} \sup_{\mathbf P\in\mathfrak P_\Delta^T}\overline\beta(\delta;\mathbf P),\qquad
  \inf_{h\in\mathcal D_\alpha}
  \sup_{P\in\mathcal P}
  \beta_{\max}(h;P),
\end{align*}
where $\mathcal D_\alpha$ is the size-constrained score class in Definition~1 and $\mathcal P$ is a
specified class of admissible source distributions. %Unlike the oracle problem, these criteria optimize against the unknown next-token laws. A uniformly most powerful pivotal score need not exist. Proving that a particular score is minimax requires a least-favorable $\mathbf P^\star$ and a saddle-point argument; minimum mean drift alone is insufficient.

\subsection{Moment-Robust Minimax-Optimal Score}
\label{sec:moment-robust-score}
We analyze detection power under the following distributional assumption.
\begin{assumption}[$\Delta$-regular distributions]
  \label{ass:Delta}
  For $0<\Delta<1/2$, let
 $\cP_\Delta
  =
  \left\{
    P:\max_{v\in\V}p(v)\leq1-\Delta
  \right\}$. % &\mathfrak P_\Delta^T =\left\{ \mathbf P:   P_t\in\cP_\Delta\ \text{for every }t\leq T\right\}.
We assume $P_t\in\cP_\Delta$ for every $t\le T$.
\end{assumption}
%Following the pivotal principle of~\cite{listatistical}, we instead restrict attention to statistics of the observable pivots $X_t$. 
Without restrictions on $P_t$, degenerate cases such as $p_t(v)=1$ make virtually all watermark
detectors ineffective. The $\Delta$-regular class excludes such cases and provides a meaningful
domain for minimax score analysis.
%The forward and reverse traversals are related by the reflection $X\mapsto-X$. To exploit this symmetry, we restrict attention to score pairs satisfying $\blue h_1(x)=h_0(-x)$. %不能是偶函数吧

%{\blue For $P\in\mathcal P_\Delta$, let $\blue Q_{0,P}^X$ denote the projected law of $X$ under forward traversal. The induced family\[  \mathcal F_\Delta^+  :=\{Q_{0,P}^X:P\in\mathcal P_\Delta\}\]is the image of $\mathcal P_\Delta$ under a nonlinear map involving random permutation, interval sampling, and projection onto $X$. It is generally difficult to characterize explicitly and need not be convex. We therefore replace $\mathcal F_\Delta^+$ by a tractable outer relaxation that retains the support and first-moment constraints shared by all of its members}. 
For a forward-traversal bit, since $P_t\in\mathcal P_\Delta$,
\eqref{eq:rank-drift} and $1-\sum_v p(v)^2\ge2\Delta(1-\Delta)$ give
\begin{align*}
  \mathbb E[X_t\mid P_t,b=0]
  \ge
  \frac{2(V+1)}{3(V-1)}
  \Delta(1-\Delta)
  := \mu_\Delta.
\end{align*}
% 要求gini?
\begin{definition}[$\mu_\Delta$-moment distribution class]
  \begin{equation*}
 \mathcal Q_\Delta
 =\left\{Q:\operatorname{supp}(Q)\subseteq[-1,1],\quad
                 \mathbb E_Q[X]\ge\mu_\Delta\right\}.
\end{equation*}
\end{definition}
The class $\mathcal Q_\Delta$ is an outer relaxation of the forward-traversal laws induced by
$\mathcal P_\Delta$: it retains only the bounded support and the guaranteed positive drift, and may
therefore contain distributions that no PIVOT channel can realize.  This relaxation makes the
minimax problem tractable and yields a uniform, conservative guarantee for every admissible
next-token law.  %Accordingly, minimax optimality over $\mathcal Q_\Delta$ should not be interpreted as exact optimality over the smaller realizable family.  
Accordingly, we refer to minimax optimality over $\mathcal Q_\Delta$ as
\emph{moment-robust minimax optimality}.
The corresponding reverse moment class is
$\mathcal Q_\Delta^-:=\{\mathcal L(-X):\mathcal L(X)\in\mathcal Q_\Delta\}$.

Under $H_0$, the pivotal property makes the conditional law of $X_t$ independent of $P_t$.
Thus, for $0<q<1$, define
\begin{equation*}
  S_V(q)
  :=
  \mathbb E\left[\frac{1}{1-qX_t}\,\middle|\,H_0\right].
\end{equation*}
There is a unique $q_{\Delta,V}\in(0,1)$ satisfying
\begin{equation}
  \bigl(1-q_{\Delta,V}\mu_\Delta\bigr)S_V(q_{\Delta,V})=1.
  \label{eq:q-delta-equation}
\end{equation}
% denote this common conditional null law by $P_0^X$ when a reference measure is needed below.
Define
\begin{align}
 &h_{0}^{\ast}(x)=-\log(1-q_{\Delta,V}x),\notag\\
 &h_{1}^{\ast}(x)=h_{0}^{\ast}(-x)
                              =-\log(1+q_{\Delta,V}x).
 \label{eq:moment-minimax-score}
\end{align}

For a forward score $h$, define its worst-case Chernoff efficiency over the moment class by
\begin{align}
 R_\Delta(h)
 &:=-\inf_{\theta\ge0}\sup_{Q\in\mathcal Q_\Delta}
 \left\{\theta\mathbb E_0[h(X)]
       +\log\mathbb E_Q[e^{-\theta h(X)}]\right\},\label{eq:moment-minimax-program}\\
 R_\Delta^\ast&:=\sup_h R_\Delta(h).
 \label{eq:moment-efficiency-rate}
\end{align}

\begin{theorem}[Moment-robust minimax-optimal score]
\label{thm:moment-minimax-score}
Under Assumption~\ref{ass:Delta}, $h_{0}^{\ast}$ attains $R_\Delta^\ast$.  %and, equivalently, solves the following minimax program: \begin{equation}\inf_h\sup_{Q\in\mathcal Q_\Delta}\left\{\mathbb E_0h(X)+\log\mathbb E_Qe^{-h(X)}\right\}.\label{eq:moment-minimax-program}\end{equation}
Let $P_0^X$ denote the distribution of $X$ under $H_0$. The least-favorable law
$Q_{\Delta,V}^\star$, defined by
\begin{equation*}
 \frac{dQ_{\Delta,V}^\star}{dP_0^X}(x)
 =\frac{1}{S_V(q_{\Delta,V})(1-q_{\Delta,V}x)},
\end{equation*}
belongs to $\mathcal Q_\Delta$, and
$(h_{0}^{\ast},Q_{\Delta,V}^\star)$ is a saddle point of
\eqref{eq:moment-minimax-program}. The optimal efficiency rate is
\begin{equation}
  R_\Delta^{\ast}
 =D_{\rm KL}(P_0^X\Vert Q_{\Delta,V}^\star).
 %=\log S_V(q_{\Delta,V}) +\mathbb E_0\log(1-q_{\Delta,V}X).
 \label{eq:moment-minimax-rate}
\end{equation}
The reflected score $h_{1}^{\ast}$ has the same guarantee for the reverse moment class
$\mathcal Q_\Delta^-$. %Common additive constants and common positive rescalings of the two scores give equivalent fixed-size tests.
\end{theorem}

%to the linear rule
Theorem~\ref{thm:moment-minimax-score} yields an explicit reciprocal-affine score and certifies its
worst-case detection rate through a saddle point.  By exploiting the full pivotal null law through
$S_V(q)$, the score upweights strongly aligned pivots relative to weakly aligned ones and can improve the
detection exponent while retaining model-free calibration.
$\mathcal Q_\Delta$ is an enlarged moment class, so the guarantee may be conservative.  Moreover, the theorem optimizes binary
detection within additive pivotal scores, not complete-message recovery, which also depends on
pairwise confusion and the codebook.
In the large-vocabulary limit, the null pivot admits an explicit product representation and
$S_V(q)$ reduces to a dilogarithm series, so the calibrating root $q_{\Delta,V}$ can be computed by a
one-dimensional root solve that does not require model access; details are given in the
supplementary material.

%\subsection{Analysis Of Test Power}
\subsection{Finite-Sample Detection and Recovery Guarantees}
%也许需要重起标题？test power

We next give a finite-sample bound for the complete-message error of PIVOT with the
score~\eqref{eq:moment-minimax-score}.  %{\blue The additional issue beyond watermark detection is pairwise message separation: the true and wrong bit scores must be ordered correctly within every block.}
For $q=q_{\Delta,V}$, define
\begin{align}
 \eta_{0,V}(q)
   &:=\mathbb E_0[-\log(1-qX)],\notag\\
 w(q)
   &:=\log\frac{1+q}{1-q}=2\operatorname{arctanh}(q),\notag\\
 \nu_{\Delta,V}
   &:=2q\mu_\Delta-
       \{\operatorname{arctanh}(q)-q\}(1-\mu_\Delta).
 \label{eq:optimal-pairwise-margin}
\end{align}
The quantity $\nu_{\Delta,V}$ is a conservative lower bound on the one-token expected score
advantage of the true bit over the wrong bit.  It is explicit once the one-dimensional
equation~\eqref{eq:q-delta-equation} has been solved.

\begin{theorem}[Complete-message power of PIVOT]
\label{thm:optimal-complete-message-power}
Suppose that the $L$ message blocks have equal length $N$, so that $T=LN$ and $m=2^L$.
Suppose further that Assumption~\ref{ass:Delta} holds, and use the scores
in~\eqref{eq:moment-minimax-score} with the threshold
\begin{equation}
 \tau_{\alpha,T}^{\ast}
 =
 T\eta_{0,V}(q)
 +w(q)\sqrt{\frac{T}{2}\log\frac{m}{\alpha}}.
 \label{eq:optimal-score-threshold}
\end{equation}
Then the decoder~\eqref{eq:generic-decoder} has $\beta_0\le\alpha$ and
\begin{align}
 \beta_{\max}(h^\ast)
 \le \min\Bigg\{1,\;&
 \exp\left[
   -T R_\Delta^\ast
   +w(q)\sqrt{\frac{T}{2}\log\frac{m}{\alpha}}
 \right]\notag\\
 &+L\exp\left[
   -\frac{N\nu_{\Delta,V}^2}{2w(q)^2}
 \right]\Bigg\}.
 \label{eq:optimal-beta-max-bound}
\end{align}
Consequently, if $\nu_{\Delta,V}>0$ and $L$ is fixed, then
$\beta_{\max}(h^\ast)\to0$ exponentially as $N\to\infty$.  The same bound holds when the exact
pivotal-null threshold is used, provided it is no larger than
$\tau_{\alpha,T}^{\ast}$.
\end{theorem}

The first term in~\eqref{eq:optimal-beta-max-bound} controls the failure of the true message score to
cross the watermark threshold and has limiting exponent $R_\Delta^\ast$.  The second controls
the selection of an incorrect bit and therefore complete-message confusion. The condition
$\nu_{\Delta,V}>0$ is sufficient rather than necessary; when it fails, the theorem is silent rather
than predicting failure.  %{\blue A corresponding bound for the linear score, which requires no additional pairwise-margin condition, is provided in the supplementary material.}

\section{Numerical Results}
\subsection{Experimental setup}
%\paragraph{Models and generation.}
We evaluate multi-bit watermarking on two publicly available models of comparable scale:
Qwen3-8B and Mistral-7B-Instruct-v0.3. We treat model identity as a fixed experimental factor
and report all metrics separately by model. Prompts are drawn from news-style English prefixes
and contain at least $50$ model tokens; each model uses a tokenizer-matched prompt pool so that
context lengths are comparable after tokenization. We use $n_{\mathrm{eval}}=128$ evaluation
prompts and a disjoint calibration set of $n_{\mathrm{calib}}=64$. Trial-level randomness
comes from the sampled prompts, uniformly drawn payloads, and, when required, independent secret
keys. Generation uses a temperature of $1.0$ without top-$k$ or top-$p$ truncation; ordinary
and watermarked continuations otherwise share the same settings.
In the experiments, we write $k=L$ for the payload length and embed uncoded payloads of
$k\in\{1,2,4\}$ bits, one bit per equal-length block.
The primary comparison uses $N{=}128$ tokens per bit ($T{=}Nk\in\{128,256,512\}$);
we additionally sweep $N\in\{16,32,64,256\}$ for PIVOT scaling
(Figure~\ref{fig:pivot-n-scaling}).
Full multi-method tables at $N\in\{64,256\}$ and calibrated PIVOT thresholds at $N{=}128$
appear in the supplementary material.
%应该只与PIVOT 有关
PIVOT decoding is blockwise and does not enumerate all $2^k$ messages.

%\paragraph{Methods and metrics.}%calibration.
We compare PIVOT, ThreeBricks-Gumbel~\cite{fernandez2023threebricks},
BiMark~\cite{feng2025bimark}, and MPAC~\cite{yoo2024multibit}.
%All three baselines are training-free, keyed multi-bit schemes with model-free verification. ThreeBricks-Gumbel uses message-dependent shifts of Gumbel-max randomness and exactly preserves the next-token law.  MPAC assigns tokens to payload positions and applies message-dependent color-list logit bias, whereas BiMark combines position allocation with XOR-masked multilayer reweighting; the former is biased and the latter is distribution-preserving only in expectation.
ThreeBricks-Gumbel is the closest exact-preservation baseline, whereas MPAC and BiMark represent
biased and expectation-unbiased position-allocation designs, respectively.
All methods use the same prompts and generation lengths.  Their detection thresholds are calibrated
separately on the unwatermarked calibration set at the nominal level $\alpha=0.05$ and frozen before
evaluation.
For PIVOT we use the moment-minimax score~\eqref{eq:moment-minimax-score} with
$\Delta=0.1$ and the large-vocabulary root $q_{\Delta,\infty}=0.5037$
(corresponding to $\mu_{\Delta,\infty}=0.06$).
%The language-model tables use PIVOT; the controlled comparison with the linear score is reported in the synthetic study in the appendix.  PIVOT verification reconstructs only keyed randomness and therefore requires no language-model query.

\textbf{Detection and recovery.}
We report the empirical false-positive rate (FPR) on ordinary text and the detection power, bit
error rate (BER), and complete-message accuracy $\widehat A_{\mathrm{msg}}$ on watermarked text.
Complete-message accuracy is our primary recovery metric: a trial is counted as correct only when
the watermark is detected and every payload bit is recovered. Hence,
$1-\widehat A_{\mathrm{msg}}$ is the empirical counterpart of the modified Type-II error.

\textbf{Distribution preservation and text quality.}
We evaluate text quality using the paired perplexity difference
\[
  \Delta\mathrm{PPL}
  =\mathrm{PPL}_{\mathrm{wm}}-\mathrm{PPL}_{\mathrm{uw}},
\]
computed under the generating model. We further probe one-step distribution preservation using two
total-variation metrics. $\mathrm{TV}_{\mathrm{mc}}$ compares the empirical token frequencies from
ordinary and watermarked sampling, whereas $\mathrm{TV}_{\mathrm{step}}$ compares the empirical
watermarked frequencies with the exact next-token distribution. For both TV metrics, we use $20$
fixed contexts and $5{,}000$ draws per context. Because these are finite-sample quantities, even a
sampler with exact distribution preservation exhibits a nonzero TV at the Monte Carlo noise floor.

\textbf{Substitution robustness.}
We additionally evaluate PIVOT under light random token substitution at rates
$5\%$, $10\%$, and $15\%$ on frozen watermarked continuations with $N{=}128$ and $k{=}2$.
For a rate-$r$ attack we replace $\mathrm{round}(rT)$ continuation tokens by
i.i.d.\ draws from the uniform distribution over the vocabulary, leave the remaining tokens and
sequence length unchanged, and decode with the clean-text threshold and the original secret key.
We report power, BER, and $\widehat A_{\mathrm{msg}}$ on the unattacked traces and at each attack rate.

\subsection{Results}
\begin{figure}[t]
  \centering
  \includegraphics[width=\columnwidth,keepaspectratio]{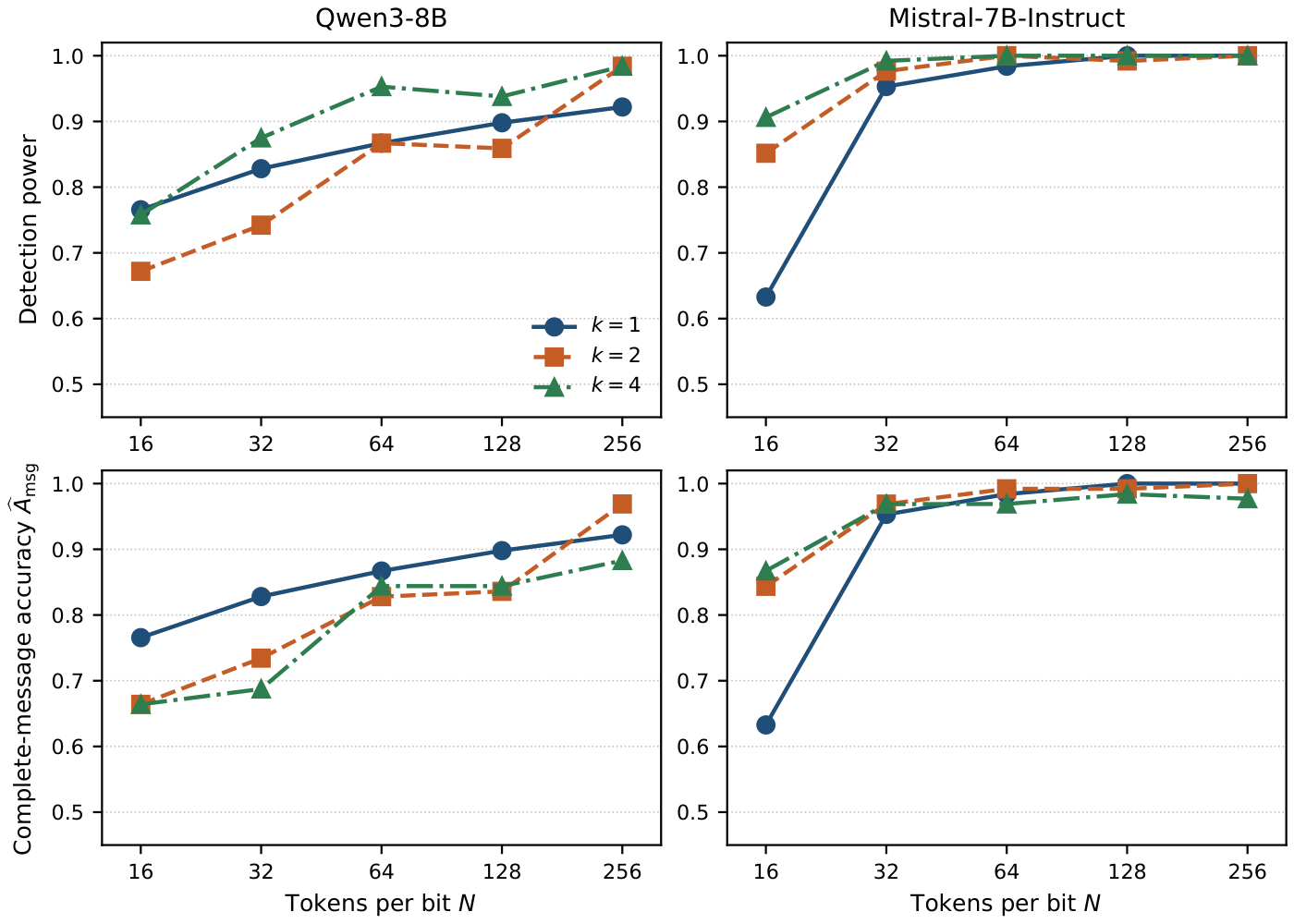}
  \caption{PIVOT detection power (top) and complete-message accuracy
  $\widehat A_{\mathrm{msg}}$ (bottom) versus tokens-per-bit $N$, for
  $k\in\{1,2,4\}$ on Qwen3-8B (left) and Mistral-7B-Instruct-v0.3 (right).
  %($T{=}Nk$, $n_{\mathrm{eval}}{=}128$, $\alpha{=}0.05$).
  }
  \label{fig:pivot-n-scaling}
  \end{figure}

Table~\ref{tab:lm-multibit} summarizes clean-text detection and payload recovery at the
primary setting $N{=}128$; Table~\ref{tab:lm-distortion} reports the three distortion probes.
Figure~\ref{fig:pivot-n-scaling} isolates PIVOT across block lengths
$N\in\{16,32,64,128,256\}$ on both models.
Table~\ref{tab:lm-attacks} reports the random-substitution probe at $N{=}128$ and $k{=}2$.
\begin{table}[t]
\centering
\setlength{\tabcolsep}{1mm}
\begin{tabular}{@{}lllcccc@{}}
\toprule
Model & $k$ & Method & FPR & Power & BER & $\widehat A$ \\
\midrule
\multirow{12}{*}{Qwen3-8B}
 & \multirow{4}{*}{1}
 & PIVOT & 0.039 & 0.898 & \textbf{0.000} & 0.898 \\
 & & ThreeBricks & 0.062 & 0.992 & 0.000 & 0.992 \\
 & & MPAC & 0.102 & 0.844 & 0.031 & 0.836 \\
 & & BiMark & 0.055 & 0.445 & 0.133 & 0.445 \\
\cmidrule(l){2-7}
 & \multirow{4}{*}{2}
 & PIVOT & 0.039 & 0.859 & \textbf{0.012} & 0.836 \\
 & & ThreeBricks & 0.039 & 0.953 & 0.012 & 0.945 \\
 & & MPAC & 0.109 & 0.961 & 0.043 & 0.906 \\
 & & BiMark & 0.070 & 0.820 & 0.039 & 0.805 \\
\cmidrule(l){2-7}
 & \multirow{4}{*}{4}
 & PIVOT & 0.086 & 0.938 & 0.047 & 0.844 \\
 & & ThreeBricks & 0.227 & 0.984 & 0.041 & 0.945 \\
 & & MPAC & 0.055 & 0.898 & 0.068 & 0.734 \\
 & & BiMark & 0.023 & 0.750 & 0.043 & 0.711 \\
\midrule
\multirow{12}{*}{\shortstack[l]{Mistral-7B\\Instruct}}
 & \multirow{4}{*}{1}
 & PIVOT & 0.125 & \textbf{1.000} & \textbf{0.000} & \textbf{1.000} \\
 & & ThreeBricks & 0.008 & 1.000 & 0.000 & 1.000 \\
 & & MPAC & 0.016 & 0.945 & 0.008 & 0.945 \\
 & & BiMark & 0.039 & 0.609 & 0.031 & 0.609 \\
\cmidrule(l){2-7}
 & \multirow{4}{*}{2}
 & PIVOT & \textbf{0.000} & 0.992 & \textbf{0.000} & 0.992 \\
 & & ThreeBricks & 0.047 & 1.000 & 0.000 & 1.000 \\
 & & MPAC & 0.062 & 1.000 & 0.000 & 1.000 \\
 & & BiMark & 0.047 & 0.852 & 0.031 & 0.844 \\
\cmidrule(l){2-7}
 & \multirow{4}{*}{4}
 & PIVOT & \textbf{0.016} & \textbf{1.000} & 0.004 & \textbf{0.984} \\
 & & ThreeBricks & 0.055 & 0.984 & 0.002 & 0.984 \\
 & & MPAC & 0.031 & 0.898 & 0.016 & 0.883 \\
 & & BiMark & 0.062 & 0.812 & 0.045 & 0.789 \\
\bottomrule
\end{tabular}
\caption{Multi-bit detection and recovery on Qwen3-8B and Mistral-7B-Instruct-v0.3
($N{=}128$, $n_{\mathrm{eval}}{=}128$, $\alpha{=}0.05$).
ThreeBricks denotes ThreeBricks-Gumbel; $\widehat A$ denotes $\widehat A_{\mathrm{msg}}$.}
\label{tab:lm-multibit}
\end{table}

\begin{table}[t]
\centering
\setlength{\tabcolsep}{1mm}
\begin{tabular}{@{}lllccc@{}}
\toprule
Model & $k$ & Method & $\Delta\mathrm{PPL}$ & TV$_{\mathrm{mc}}$ & TV$_{\mathrm{step}}$ \\
\midrule
\multirow{12}{*}{Qwen3-8B}
 & \multirow{4}{*}{1}
 & PIVOT & $-0.65$ & 0.064 & 0.054 \\
 & & ThreeBricks & $-0.66$ & 0.064 & 0.052 \\
 & & MPAC & $+0.24$ & 0.353 & 0.351 \\
 & & BiMark & $+0.14$ & 0.419 & 0.417 \\
\cmidrule(l){2-6}
 & \multirow{4}{*}{2}
 & PIVOT & $-0.20$ & 0.067 & 0.052 \\
 & & ThreeBricks & $-1.11$ & 0.064 & 0.051 \\
 & & MPAC & $+1.05$ & 0.338 & 0.333 \\
 & & BiMark & $+0.48$ & 0.438 & 0.435 \\
\cmidrule(l){2-6}
 & \multirow{4}{*}{4}
 & PIVOT & $-0.82$ & 0.066 & 0.052 \\
 & & ThreeBricks & $-0.52$ & 0.065 & 0.052 \\
 & & MPAC & $+0.11$ & 0.319 & 0.319 \\
 & & BiMark & $-0.13$ & 0.429 & 0.428 \\
\midrule
\multirow{12}{*}{\shortstack[l]{Mistral-7B\\Instruct}}
 & \multirow{4}{*}{1}
 & PIVOT & $+1.17$ & 0.048 & 0.039 \\
 & & ThreeBricks & $-2.30$ & 0.051 & 0.040 \\
 & & MPAC & $+0.20$ & 0.264 & 0.259 \\
 & & BiMark & $-0.82$ & 0.342 & 0.338 \\
\cmidrule(l){2-6}
 & \multirow{4}{*}{2}
 & PIVOT & $-0.85$ & 0.049 & 0.040 \\
 & & ThreeBricks & $-2.40$ & 0.050 & 0.039 \\
 & & MPAC & $-2.33$ & 0.282 & 0.281 \\
 & & BiMark & $-0.90$ & 0.343 & 0.342 \\
\cmidrule(l){2-6}
 & \multirow{4}{*}{4}
 & PIVOT & $-1.77$ & 0.048 & 0.037 \\
 & & ThreeBricks & $-3.07$ & 0.051 & 0.040 \\
 & & MPAC & $-1.93$ & 0.254 & 0.256 \\
 & & BiMark & $-0.44$ & 0.359 & 0.357 \\
\bottomrule
\end{tabular}
\caption{Distribution-preservation probes on Qwen3-8B and Mistral-7B-Instruct-v0.3
(means over $20$ contexts, with $5{,}000$ draws per context).
ThreeBricks denotes ThreeBricks-Gumbel.}
\label{tab:lm-distortion}
\end{table}

\iffalse
\begin{table}[t]
\centering
\setlength{\tabcolsep}{2.4pt}
\begin{tabular}{@{}c@{\hspace{5pt}}ccc@{\hspace{5pt}}ccc@{}}
\toprule
\multicolumn{7}{c}{Qwen3-8B} \\
\cmidrule(lr){1-7}
$N$ & \multicolumn{3}{c}{Power} & \multicolumn{3}{c}{$\widehat A_{\mathrm{msg}}$} \\
\cmidrule(lr){2-4}\cmidrule(lr){5-7}
 & $k{=}1$ & $k{=}2$ & $k{=}4$ & $k{=}1$ & $k{=}2$ & $k{=}4$ \\
\midrule
64  & 0.867 & 0.867 & 0.953 & 0.867 & 0.828 & 0.844 \\
128 & 0.898 & 0.859 & 0.938 & 0.898 & 0.836 & 0.844 \\
256 & 0.922 & 0.984 & 0.984 & 0.922 & 0.969 & 0.883 \\
\midrule
\multicolumn{7}{c}{Mistral-7B-Instruct-v0.3} \\
\cmidrule(lr){1-7}
$N$ & \multicolumn{3}{c}{Power} & \multicolumn{3}{c}{$\widehat A_{\mathrm{msg}}$} \\
\cmidrule(lr){2-4}\cmidrule(lr){5-7}
 & $k{=}1$ & $k{=}2$ & $k{=}4$ & $k{=}1$ & $k{=}2$ & $k{=}4$ \\
\midrule
64  & 0.984 & 1.000 & 1.000 & 0.984 & 0.992 & 0.969 \\
128 & 1.000 & 0.992 & 1.000 & 1.000 & 0.992 & 0.984 \\
256 & 1.000 & 1.000 & 1.000 & 1.000 & 1.000 & 0.977 \\
\bottomrule
\end{tabular}
\caption{PIVOT versus block length $N$
on Qwen3-8B and Mistral-7B-Instruct-v0.3.
%($T{=}Nk$, $n_{\mathrm{eval}}{=}128$, $\alpha{=}0.05$).
}
\label{tab:pivot-n-scaling}
\end{table}
\fi
%\paragraph{Analysis.}
Several patterns emerge from Tables~\ref{tab:lm-multibit}--\ref{tab:lm-attacks}
and Figure~\ref{fig:pivot-n-scaling}.
\begin{enumerate}
\item \textbf{Recovery vs.\ Type-I control.}
On Qwen3-8B, ThreeBricks-Gumbel attains the highest $\widehat A_{\mathrm{msg}}$ at every $k$, with
near-perfect recovery at $k=1$.  However, at $k=4$ its empirical FPR rises to $0.227$, far above the
nominal $\alpha=0.05$, so the comparison no longer reflects the nominal Type-I level.
PIVOT provides a more balanced operating point: at $k=4$ it has FPR $0.086$ with
$\widehat A_{\mathrm{msg}}=0.844$ and power $0.938$.
On Mistral-7B-Instruct-v0.3, the same trade-off is less pronounced: PIVOT matches or nearly matches the
best recovery %($\widehat A_{\mathrm{msg}}=1.000/0.992/0.984$ for $k=1/2/4$) 
while keeping FPR at or
below the nominal level for $k\!\ge\!2$ (FPR $0.000$ and $0.016$), whereas ThreeBricks-Gumbel remains
near $\alpha$ at larger $k$.
\item \textbf{Distortion aligns with theory.}
Across both models, PIVOT and ThreeBricks-Gumbel exhibit
$\mathrm{TV}_{\mathrm{mc}}\!\approx\!0.05$--$0.07$, close to the Monte Carlo noise floor for
exact distribution-preserving sampling.
MPAC and BiMark incur large one-step TVs ($\approx0.25$--$0.44$), consistent with their biased and
expectation-unbiased constructions, respectively.
Paired $\Delta\mathrm{PPL}$ varies across settings and should not be over-interpreted; the TV
probes provide more stable distortion summaries.
\item \textbf{Payload scaling at fixed $N$.}
Holding tokens-per-bit fixed at $N=128$, increasing $k$ lengthens $T$ but also multiplies the
number of bits that must be jointly correct. On Qwen3-8B, BER rises gently with $k$ for
PIVOT; on Mistral it stays near zero.
BiMark remains the weakest recovery method at $k=1$ on both models and never leads on
$\widehat A_{\mathrm{msg}}$. MPAC is competitive at $k=2$ but degrades at $k=4$ on Qwen3-8B
and, more mildly, on Mistral, indicating detection without fully reliable decoding as the
payload grows.
\item \textbf{Scaling with block length $N$.}
Figure~\ref{fig:pivot-n-scaling} shows that increasing tokens-per-bit generally strengthens
PIVOT recovery, with the largest gains at short blocks.
On Qwen3-8B the trend is steady across $k$, though at larger payloads detection can improve
faster than complete-message recovery.
On Mistral-7B-Instruct-v0.3, only the shortest blocks remain weak; once $N$ is moderate,
recovery is near ceiling for all $k$.
%\item \textbf{Cross-model consistency.}The qualitative ranking is stable: exact-preservation methods uniquely combine strong recovery with near-zero token-level distortion, while MPAC and BiMark exhibit substantially larger TV deviations. Absolute operating points differ---Mistral yields higher PIVOT recovery and lower FPR at $k\!\ge\!2$ than Qwen3-8B under the same $N$---but the conclusion is unchanged.
\item
\textbf{Substitution robustness.}
Table~\ref{tab:lm-attacks} shows that under $5\%$, $10\%$, and $15\%$ random token replacement,
power and $\widehat A_{\mathrm{msg}}$ degrade only gradually relative to the unattacked baseline
on the same traces: on Mistral they remain near ceiling, and on Qwen3-8B
$\widehat A_{\mathrm{msg}}$ falls from $0.930$ to $0.844$ at $15\%$, indicating that the keyed
pivot signal is not immediately erased by scattered token flips.
\end{enumerate}
\iffalse
\begin{table}[t]
  \centering
  \setlength{\tabcolsep}{3.5pt}
  \begin{tabular}{@{}llcccc@{}}
  \toprule
  Model & Attack & Power & BER & $\widehat A_{\mathrm{msg}}$ \\
  \midrule
  \multirow{3}{*}{\shortstack[l]{Qwen3-8B\\($k{=}1$)}}
   & Clean & 0.898 & 0.008 & 0.898 \\
   & Sub.\ $5\%$ & 0.891 & 0.016 & 0.891 \\
   & Sub.\ $10\%$ & 0.875 & 0.023 & 0.875 \\
  \midrule
  \multirow{3}{*}{\shortstack[l]{Mistral-7B\\Instruct-v0.3\\($k{=}2$)}}
   & Clean & 1.000 & 0.000 & 1.000 \\
   & Sub.\ $5\%$ & 0.977 & 0.000 & 0.977 \\
   & Sub.\ $10\%$ & 0.992 & 0.000 & 0.992 \\
  \bottomrule
  \end{tabular}
  \caption{Light random-substitution robustness of PIVOT on frozen watermarked traces.
  %($N{=}128$, $n{=}128$; thresholds fixed from the clean Type-I calibration).
  Each attacked token is replaced by a uniform random vocabulary id.}
  \label{tab:lm-attacks}
  \end{table}
\fi
\begin{table}[htpb]
  \centering
  \setlength{\tabcolsep}{3.5pt}
  \begin{tabular}{@{}llccc@{}}
  \toprule
  Model & Attack & Power & BER & $\widehat A_{\mathrm{msg}}$ \\
  \midrule
  \multirow{4}{*}{\shortstack[l]{Qwen3-8B}}
   & Clean & 0.953 & 0.012 & 0.930 \\
   & Sub.\ $5\%$ & 0.953 & 0.023 & 0.914 \\
   & Sub.\ $10\%$ & 0.922 & 0.027 & 0.875 \\
   & Sub.\ $15\%$ & 0.883 & 0.027 & 0.844 \\
  \midrule
  \multirow{4}{*}{\shortstack[l]{Mistral-7B\\Instruct-v0.3}}
   & Clean & 0.992 & 0.000 & 0.992 \\
   & Sub.\ $5\%$ & 0.992 & 0.000 & 0.992 \\
   & Sub.\ $10\%$ & 0.977 & 0.000 & 0.977 \\
   & Sub.\ $15\%$ & 0.961 & 0.000 & 0.961 \\
  \bottomrule
  \end{tabular}
  \caption{Light random-substitution robustness of PIVOT on frozen watermarked traces
  ($N{=}128$, $k{=}2$).}
  \label{tab:lm-attacks}
\end{table}
\section{Conclusion}
\label{sec:conclusion}

PIVOT combines exact message-conditional distribution preservation, model-free calibration, and
moment-robust score design.  Its ``optimality'' concerns binary Chernoff efficiency for additive
pivotal scores over the relaxed class $\mathcal Q_\Delta$, not global complete-message optimality
or robustness to arbitrary text edits. %Future work should characterize realizable alternatives, jointly optimize scores and codebooks, and evaluate longer payloads, robustness to attacks, and a broader range of models.
%Future work should characterize realizable alternatives, jointly optimize scores and codebooks, and evaluate longer payloads, robustness to attacks, and a broader range of models.

\bibliography{references}

\end{document}